\documentclass[conference]{IEEEtran}
\IEEEoverridecommandlockouts

\usepackage{amssymb,upgreek,nicefrac,amsmath}
\usepackage[usenames,dvipsnames]{xcolor}
\usepackage[dvips]{graphicx}
\usepackage{listings}
\usepackage{multirow}
\usepackage{mdframed}
\usepackage{hyperref}
\usepackage{soul}

\usepackage{algorithm}
\usepackage[noend]{algpseudocode}
\usepackage{xtab,afterpage,lipsum}
\usepackage{bm}

\definecolor{Yellow}{rgb}{1, 1, 0}
\definecolor{VeryLightGray}{gray}{.90}
\definecolor{LightGray}{gray}{.7}
\definecolor{Gray}{gray}{.50}
\definecolor{DarkGray}{gray}{.3}
\definecolor{VeryDarkGray}{gray}{.10}

\newcommand{\magn}[1]{\left\vert #1 \right\vert}

\newcommand{\df}{\doteq}

\newcommand{\RM}{$\mathsf{RM}$}
\newcommand{\RL}{$\mathsf{PaRLSched}$}

\newcommand{\OS}{$\mathsf{OS}$}

\newcommand{\RAND}[2]{{\sf rand}_{#1}\left[#2\right]}
\newcommand{\cI}{\mathcal{I}}

\newtheorem{proposition}{Proposition}[section]

\newenvironment{proof}{\textbf{Proof.}}{$\square$\\}

\begin{document}

\title{Learning-based Dynamic Pinning of Parallelized Applications in Many-Core Systems \thanks{This paper is an extension of an earlier version appeared in the conference paper \cite{chasparis_learning-based_2019}. It has been supported by the European Union grant EU H2020-ICT-2014-1 project RePhrase (No. 644235).}}

\author{Georgios C. Chasparis\thanks{G. C. Chasparis and M. Rossbory are with the Software Competence Center Hagenberg GmbH, Softwarepark 21, A-4232 Hagenberg, Austria.} \and Vladimir Janjic\thanks{V. Janjic is with the School of Computer Science, University of St Andrews, Scotland, UK.} \and Michael Rossbory
}


\maketitle

\begin{abstract}
Motivated by the need for adaptive, secure and responsive scheduling in a great range of computing applications, including human-centered and time-critical applications, this paper proposes a scheduling framework that seamlessly adds resource-awareness to any parallel application. In particular, we introduce a learning-based framework for dynamic placement of parallel threads to Non-Uniform Memory Access (NUMA) architectures. Decisions are taken independently by each thread in a decentralized fashion that significantly reduces computational complexity. The advantage of the proposed learning scheme is the ability to easily incorporate any multi-objective criterion and easily adapt to performance variations during runtime. Under the multi-objective criterion of maximizing total completed instructions per second (i.e., both computational and memory-access instructions), we provide analytical guarantees with respect to the expected performance of the parallel application. We also compare the performance of the proposed scheme with the Linux operating system scheduler in an extensive set of applications, including both computationally and memory intensive ones. We have observed that performance improvement could be significant especially under limited availability of resources and under irregular memory-access patterns.
\end{abstract}


\section{Introduction}

\noindent

Efficient resource allocation for multi-threaded applications in NUMA architectures has attracted significant scientific attention due to a) the involved \emph{complexity} of the decision-making process, and b) the need to incorporate \emph{alternative optimization criteria} that goes beyond standard maximization of execution speed. This statement is further reinforced by the recent advancement of tools for parallelizing complex applications, that gave birth to non-trivial and highly advanced parallel and data patterns \cite{Danelutto2001,Aldinucci2015_PVD,delrioastorga_generic_2017,RPL}. In addition, the nature of an application (e.g., machine-learning, image processing, control and optimization) may add additional criteria that cannot easily be integrated into an OS scheduler. As expected, \emph{the problem of efficiently utilizing resources, while concurrently optimizing a multi-objective criterion, cannot be treated by standard heuristic-based techniques.}

To this end, this paper proposes and investigates the potential of a learning- or measurement-based scheduling scheme that is part of a running application and regularly corrects/improves allocation decisions given the observed application's performance.
%
In particular, this paper proposes a distributed learning scheme specifically tailored for addressing the problem of dynamically assigning/pinning threads of a parallelized application to the available processing units. The proposed scheme is flexible enough to incorporate any multi-objective optimization criterion and provides convergence guarantees to at least suboptimal assignments. Given the fact that it is measurement-based, it is computationally efficient with a linear-complexity with the number of threads. Since it is iterative in nature, it also exhibits minimal memory requirements. 

It is worth noting that we target an \emph{online} learning framework where allocation decisions are taken during runtime, and without requiring any prior application knowledge. Such feature can make parallel applications more responsive by reducing their execution time, especially in situations where computing resources are shared between different applications. This is also very important for human-centered computing, where strict timing requirements can be of high importance, given that they are often computationally intensive, such as machine-learning or image processing applications. In addition, the proposed scheduling framework can seamlessly be attached to any parallel application. These features provide an easy-to-use and user-friendly supervisory scheduling scheme that reduces the need for expert and application knowledge.


In our previous work~\cite{chasparis_euro-par_2017,chasparis_efficient_2017}, we have proposed a reinforcement-learning-based distributed scheduling framework (\RL), adapted to Uniform Memory Architectures (UMA). In this paper, our goal is to provide a generalized methodology that also extends to Non-Uniform Memory Architectures (NUMA). Such framework should be considered as a supervisory scheme that acts on top of any OS scheduling and performs either low- or high-frequency allocation corrections possibly subject to alternative multi-objective criteria. For example, when optimizing with respect to both computational and memory-access instructions completed  per second, the learning scheme should find the right balance between computing bandwidth and memory affinities. In this paper though, we are not concerned with memory migrations.

This paper is an extension of an earlier version appeared in \cite{chasparis_learning-based_2019}. In this updated version, we provide analytical guarantees of the performance of the learning-based scheduling framework, and we have extended our experimental evaluation to applications with memory irregularities. 

The paper is organized as follows. Section~\ref{sec:RelatedWork} discusses related work and contributions. Section~\ref{sec:framework} describes the problem formulation and objective of the paper. Section~\ref{sec:DynamicScheduler} presents the main features of the proposed Dynamic Scheduler (\RL) and Section~\ref{sec:ConvergenceAnalysis} provides analytical convergence guarantees with respect to the application's performance. Section \ref{sec:Experiments} presents a performance comparison with the standard Linux scheduler in benchmark applications. Finally, Section~\ref{sec:Conclusions} presents concluding remarks and future work.

\section{Related Work and Contributions}	\label{sec:RelatedWork}

\noindent
Prior work has demonstrated the importance of thread-to-core bindings in the overall performance of a parallelized application \cite{podzimek_analyzing_2015}. The task of discovering such optimal bindings is rather complex, given the structure of NUMA architectures \cite{goglin_managing_2014}. This task becomes even harder given the need for developing tools that can easily generalize to any architecture and they are application independent. 

For example, reference \cite{Klug11} describes a tool that checks the performance of each of the available thread-to-core bindings and searches for an optimal placement. Unfortunately, the \emph{exhaustive-search} type of optimization that is implemented may prohibit runtime implementation. Reference~\cite{Broquedis10}  combines the problem of thread scheduling with \emph{scheduling hints} related to thread-memory affinity issues. A similar scheduling policy is also implemented by \cite{Olivier11}. 

At the same time, given that no prior knowledge of the application's details is available, a centralized optimization formulation is prohibitive. Such design restrictions give rise to learning-based techniques, where scheduling decisions are taken based only on performance measurements. This need for learning from data has been recognized in \cite{castro_machine_2011}, where a machine learning based mechanism is designed for transactional applications. In this case, each instance of the application has to be run and profiled before any learning process is to be implemented. 

Even such learning processes could be computationally complex given the quite large search space. For this reason, distributed or game-theoretic optimizations have been attempted in the past for related problems, including cooperative game formulation for allocating bandwidth in grid computing \cite{Sub08}, the non-cooperative game formulation in the problem of medium access protocols in communications \cite{Tembine09} or for allocating resources in cloud computing \cite{Wei10}. These approaches can significantly reduce the involved computational complexity and also allow for the development of online selection rules based on performance measurements. However, such modeling techniques have not yet been implemented in the context of pinning of parallelized applications.

Recognizing this need for both learning- and distributed-based optimization, and contrary to the aforementioned references on pinning of parallelized applications, our earlier work~\cite{chasparis_euro-par_2017,chasparis_efficient_2017} proposed a scheduling scheme for optimally allocating threads of a parallelized application that combines both a learning- and a distributed-based optimization. It requires a minimum information exchange, where only measurements collected from each running thread are needed. Furthermore, it is flexible enough to accommodate alternative optimization criteria depending on the available performance counters. 
However, one potential drawback was the fact that no special consideration was taken upon the possible \emph{non-uniform memory access} (NUMA) architectures, as it did not distinguish between moving a thread to a ``local'' (within the same NUMA node) and ``remote'' (from a different NUMA node) core. 

This paper extends the scheduling framework of our previous work \cite{chasparis_euro-par_2017,chasparis_efficient_2017} with respect to the following contributions:
\begin{itemize}
\item[(C1)] We propose a novel two-level scheduling process that is appropriate for NUMA architectures. At the higher level, the scheduler decides on which NUMA node each thread should be assigned, while at the lower level it decides on which CPU core (within that NUMA node) to execute the thread.
\item[(C2)] We provide analytical convergence guarantees with respect to the resulting performance of the application in comparison to the optimal performance.
\item[(C3)] We demonstrate the efficiency of the proposed approach on several benchmark applications with different characteristics, including computational- and memory-intensive applications.
\end{itemize}
This paper is also an extension of an earlier version appeared in \cite{chasparis_learning-based_2019} with respect to contributions (C2) and (C3).

\section{Problem Formulation and Objective} 
\label{sec:framework}


Let a parallel application comprise $n$ threads, $\mathcal{I}=\{1,2,...,n\}$. We denote the \emph{assignment} of a thread $i$ to a set of available NUMA nodes $\mathcal{J}_{\rm NUMA}$ by $\alpha_i \in \mathcal{J}_{\rm NUMA}$. Within the selected NUMA node $\alpha_i$, thread $i$ should be assigned to one of the available CPU cores $\mathcal{J}_{\rm CPU}(\alpha_i)$, denoted by $\beta_i \in\mathcal{J}_{\rm CPU}(\alpha_i)$. Let also $\alpha=\{(\alpha_i,\beta_i)\,, i \in \mathcal{I}\}$ denote the overall \emph{assignment profile}, and let $\mathcal{A}$ be the set of all profiles.

The Resource Manager (\RM) periodically checks the performance of a thread and makes decisions about its assignment for the next scheduling iteration. \emph{\textbf{For the remainder of the paper}}, we will assume that: a) The internal properties and details of the threads are not known to the \RM{}. Instead, the \RM\ may only have access to measurements related to their performances; b) Threads may not be idled or postponed by the \RM. Instead, the goal of the \RM{} is to assign the \emph{currently} available resources to the \emph{currently} running threads (\emph{work-conserving}). 

\subsubsection{Static optimization and issues} 	\label{sec:StaticOptimization}

A possible centralized objective that we may consider could be to maximize the average processing speed over all threads, i.e.,
\begin{eqnarray} \label{eq:CentralizedObjective}
\max_{\alpha\in\mathcal{A}} & f(\alpha,w)\df \sum_{i=1}^{n} u_i(\alpha,w)/n,
\end{eqnarray}
where, for example, $u_i$ may represent the processing speed of thread $i$ under assignment $\alpha\in\mathcal{A}$. In general, $u_i$ will depend on the assignment profile $\alpha$ and exogenous disturbances (e.g., other applications) summarized within the parameter $w$. Any solution to the optimization problem (\ref{eq:CentralizedObjective}) will correspond to an \emph{efficient/optimal assignment}.  However, there are two practical issues when posing an optimization problem in this form, namely a) the details of the function $u_i(\alpha,w)$ are unknown and it may only be evaluated through measurements, denoted by $\tilde{u}_i$; and, b) $w$ is also unknown and may vary with time.

\subsubsection{Measurement- or learning-based optimization}
\label{sec:MeasurementBasedOptimization}

We wish to address a \emph{static} optimization objective of the form (\ref{eq:CentralizedObjective}) through a \emph{measurement-} or \emph{learning-based} methodology. That is, the \RM\ reacts to measurements of $f(\alpha,w)$, periodically collected at time instances $k=1,2,...$ and denoted by $\tilde{f}(k)$. The measured objective may take on the form $\tilde{f}(k)\df\sum_{i=1}^{n}\tilde{u}_i(k)/n$. Given these measurements and the current assignment $\alpha(k)$ of resources, the \RM\ will select the next assignment of resources $\alpha(k+1)$, so that the measured objective approaches the true optimum of the unknown performance function $f(\alpha,w)$. 




\subsubsection{Multi-agent formulation}


We further \emph{distribute} the decision-making process into a thread-based optimization, where the \RM\ makes decisions \emph{independently} for each thread. Equivalently, we may assume that each thread makes its own independent decisions as in multi-agent formulations. Such distribution reduces the complexity of the decision-making process, since each thread has a reduced number of choices as compared to the number of choices of the group of threads. Furthermore, it increases robustness, since any performance degradation noticed in a group of threads can immediately be treated by the affected threads, thus avoiding the complexity of centrally designed assignment corrections.

\subsubsection{Multi-level decision-making and actuation}	\label{sec:MultipleLevelDecisionMakingAndActuation}

Recent work by the authors \cite{chasparis_euro-par_2017,chasparis_efficient_2017} has demonstrated the potential of learning-based optimization in UMA architectures. However, when an application runs on a NUMA architecture, additional information can be exploited to enhance scheduling of a parallelized application. 
To this end, a multi-level decision-making and actuation process is considered. 
We extend the \RL\ dynamic scheduler of \cite{chasparis_euro-par_2017,chasparis_efficient_2017} by introducing two nested decision processes depicted in Figure~\ref{fig:TwoLevelSched}. At the \emph{higher level} (Level~1), the performance of a thread is evaluated with respect to its own prior history of performances, and decisions are taken with respect to its NUMA placement. At the \emph{lower level} (Level~2), the performance of a thread is evaluated with respect to its own prior history of performances, and decisions are taken with respect to its CPU placement (within the selected NUMA node).

\section{Dynamic Scheduler}		\label{sec:DynamicScheduler}

\begin{figure}
\centering
\includegraphics[scale=1]{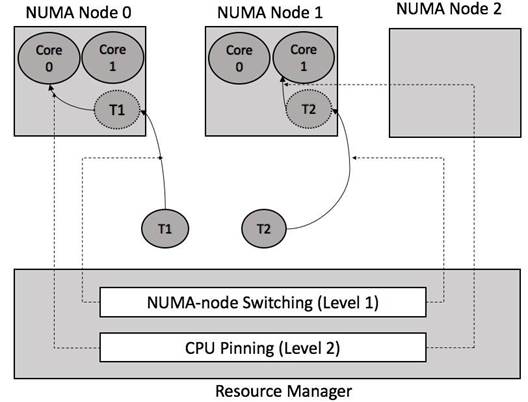}
\caption{Two-level scheduling where the \RM\ decides firstly the NUMA node and secondly the CPU core at which each thread should be pinned on.}
\label{fig:TwoLevelSched}
\end{figure}

Each one of the two levels of the decision process will take place at different frequencies and based on different reasoning. In particular, NUMA-node switching may be costly, especially when performed with high frequency due primarily to memory affinities, while CPU-node switching within the same NUMA node may be costless (with respect to its impact to the processing speed). For this reason, we have introduced two measurement-based learning algorithms specifically tailored to accommodate these different needs (Figure~\ref{fig:TwoLevelSched}):
\begin{itemize}
\item \emph{\textbf{(Level 1) Aspiration learning for NUMA-node switching}}, that responds only to significant performance variations and does not require frequent migrations. 
\item \emph{\textbf{(Level 2) Perturbed learning automata for CPU-core pinning}} within a given NUMA node, that allows frequent CPU-core switches.
\end{itemize}

We introduce periodic time instances with period $T_{\rm CPU}>0$, and indexed by $k=1,2,...$, at which decisions at Level 2 (CPU-core pinning) are revised. Decisions at Level 1 (NUMA-node switching) are performed less frequently, at periodic time instances of period $T_{\rm NUMA} \gg T_{\rm CPU}$, which will be indexed by $\tau= 1,2,...$. 

\subsection{Utility Function}

A cornerstone in the design of any such multi-agent formulation is the \emph{preference criterion} or \emph{utility function} $u_i$ for each thread $i\in\mathcal{A}$. The utility function captures the benefit of a decision maker (thread) resulting from the assignment profile $\alpha$, i.e., it represents a function of the form $u_i:\mathcal{A}\to\mathbb{R}_+$ (where we restrict it to be a positive number). The action profile (i.e., the selections of all threads) constitutes a ``state'' of the environment that directly determines the performances of all threads. We are interested in building \emph{learning-based reflex} agents that respond only to current measurements in an effort to ``eventually'' learn to play efficient assignments.

It is important to note that the utility function $u_i$ of each agent/thread $i$ is subject to \emph{design} and it is introduced in order to guide the preferences of each agent. Thus, $u_i$ may not necessarily correspond to a measured quantity, but it could be a function of available performance counters. For example, a natural choice for the utility of each thread is its own execution speed, which can be measured by the number of executed instructions per unit of time. This may also be combined with other counters, e.g., the number of memory-access instructions, the number of cache misses, etc., to give a better representation of the performance of a thread. 


\subsection{Aspiration learning for NUMA-node switching}	 \label{sec:AspirationLearning}

\noindent
We developed a novel learning scheme for NUMA-node switching that is based upon the notions of benchmark actions/performances and bears similarities with the so-called \emph{aspiration learning} \cite{ChasparisAriShamma13_SIAM}. The novelty here lies in the introduction of two benchmark levels in order to handle the possibility of noisy measurements. Such type of learning dynamics tries to gradually reach assignment profiles where all threads perform well. They have the advantage that exploration (of new assignments) can be performed selectively (e.g., when a significant reduction in performance is observed). In this way, a low-frequency NUMA-node switching can be attained. The specific steps are depicted in Table~\ref{Tb:AspirationLearningForNUMAPlacement}.

\begin{table}[t!]
\caption{Aspiration Learning for NUMA-node Switching}
\boxed{
\begin{minipage}{0.48\textwidth}
At fixed periodic time instances denoted by $\tau=1,2,...$, with period $T_{\rm NUMA}$ sec, the following steps are executed recursively for each thread $i$ in parallel.

(1) {\bf Performance measurement.} For the currently selected NUMA-node $\alpha_i(\tau)$ thread $i$ retrieves its current performance measurement, $\tilde{u}_i(\tau)$.

(2) {\bf Aspiration-level update.} Given the current performance measurement $\tilde{u}_i(\tau)$, update the discounted running average performance of the thread, as follows:
\begin{equation}	\label{eq:AspirationLevelUpdateRule}
\rho_i(\tau+1) = \rho_i(\tau) + \nu \cdot [ \tilde{u}_i(\tau) - \rho_i(\tau) ],
\end{equation}
where $\tilde{u}_i(\tau)$ is the current measurement of the utility of thread $i$.

(3) {\bf Benchmarks update.} Define the \emph{upper benchmark performance}, $\overline{b}_i(\tau)$, as a performance threshold over which a performance is considered \emph{satisfactory}, and the \emph{lower benchmark performance}, $\underline{b}_i(\tau)$, as a performance threshold under which a performance is considered \emph{unsatisfactory}, with $\underline{b}_i(\tau) < \overline{b}_i(\tau)$. They are updated as follows: 
\begin{itemize}
\item if $\rho_i(\tau+1) \geq \overline{b}_i(\tau)$, then
\begin{align*}
\overline{b}_i(\tau+1) & = \rho_i(\tau+1) \\
\underline{b}_i(\tau+1) & = \rho_i(\tau+1) / \eta 
\end{align*}
\item if $\underline{b}_i(\tau) \leq \rho_i(\tau+1) < \overline{b}_i(\tau)$, then
\begin{align*}
\overline{b}_i(\tau+1) & = \overline{b}_i(\tau)\\
\underline{b}_i(\tau+1) & = \underline{b}_i(\tau)
\end{align*}
\item if $\rho_i(\tau+1) < \underline{b}_i(\tau)$, then
\begin{align*}
\overline{b}_i(\tau+1) & = \eta\cdot \rho_i(\tau+1) \\
\underline{b}_i(\tau+1) & = \rho_i(\tau+1)
\end{align*}
\end{itemize}
for some constant $\eta > 1$.

(4) {\bf Action update.} A thread $i$ selects actions according to the following rule:
\begin{enumerate}
 \item[a)] if $\rho_i(\tau+1) < \underline{b}_i(\tau)$, i.e., if the updated discounted running average performance is unsatisfactory, then thread $i$ will perform a random switch to a better reply, i.e.,
 $$\alpha_i(\tau+1) \in {\rm rand}_{\rm unif}\left[{\rm BR}_{{\rm NUMA},i}(\alpha)\right],$$
where ${\rm BR}_{{\rm NUMA},i}(\alpha)$ denotes the better-reply of thread $i$ to the assignment $\alpha$, defined as
\begin{eqnarray}    \label{eq:NUMABetterReplyCondition}
\lefteqn{{\rm BR}_{{\rm NUMA},i}(\alpha) \df } \cr && \left\{\alpha_i'\in\mathcal{J}_{\rm NUMA}: \rho_i(\tau) < \gamma \frac{\sum_{\{j\in\mathcal{I}:\alpha_j(\tau)=\alpha_i'\}}\rho_j(\tau)}{\magn{\{j\in\mathcal{I}:\alpha_j(\tau)=\alpha_i'\}}} \right\}
\end{eqnarray}
for some $\gamma\in(0,1)$. The set $\{j\in\mathcal{I}:\alpha_j(\tau-1)=\alpha_i'\}$ includes all those threads that selected action $\alpha_i'$ in the previous time instance. In other words, an action $\alpha_i'\in{\rm BR}_{{\rm NUMA},i}(\alpha)$ if the average of the threads selecting $\alpha_i'$ did better on average than thread $i$.

If more than one thread has chosen to migrate, then only one thread (selected at random) is allowed to execute this migration.

 \item[b)] if $\rho_i(\tau+1) \geq \underline{b}_i(\tau)$ , then each thread $i$ will keep playing the same action with high probability and experiment with any other action with a small probability $\zeta>0$, i.e.,
 \begin{eqnarray}   \label{eq:AspirationLevelExperimentation}
 \alpha_i(\tau+1) = \begin{cases}
 \alpha_i(\tau),& \mbox{w.p. } 1-\zeta\\
 {\rm rand}_{\rm unif}[{\rm BR}_{{\rm NUMA},i}(\alpha)], & \mbox{w.p. } \zeta
 \end{cases}
 \end{eqnarray}

 If more than one thread has chosen to migrate, then only one thread (selected at random) is allowed to execute this migration.
\end{enumerate}
\end{minipage}
}
\label{Tb:AspirationLearningForNUMAPlacement}
\end{table}

It is important to note that this learning scheme will react immediately to a rapid drop in the performance. In particular, when the performance drops below the lower benchmark, then with high probability the action will change, while in any other case, the action will change with a small probability $\zeta>0$. The reason for maintaining both an upper and lower benchmark is in order to minimize the effect of noise in the decision-making process. 

When the thread needs to select a new NUMA node, it will select among the set of better replies, i.e., nodes at which other threads perform better so far. Note that a thread may not have a-priori knowledge of the exact impact an action switch has on his own utility (until this action switch is performed). However, we may use prior data of the performances of other threads, as defined in ${\rm BR}_{{\rm NUMA},i}(\alpha)$. Thus, at step (4a), we may direct threads that currently do not perform well to the NUMA nodes where threads perform better. 


\subsection{Perturbed Learning Automata for CPU-core pinning} \label{sec:PerturbedLearningAutomata}

Let us assume that, at Level~1, and for each one of the running threads $i\in\mathcal{I}$, the \RM\ has already selected a NUMA node $\alpha_i\in\mathcal{J}_{\rm NUMA}$. Then, at Level~2, the \RM\ needs to decide which CPU core each thread should be pinned to. Given that CPU-core switching within the same NUMA node is usually costless, we have designed a learning algorithm that allows frequent switching and therefore a faster convergence rate. To this end, we employ \emph{perturbed learning automata} \cite{chasparis_stochastic_2019} developed by the authors. Such dynamics perform well in the presence of noise contrary to alternative schemes, as discussed in \cite{chasparis_stochastic_2019}, and can guarantee convergence to at least locally optimal assignments.

The basic idea behind learning automata is rather simple. Each agent $i$ keeps track of a strategy vector that holds its estimates over the best choice. We denote this strategy by $\sigma_i=[\sigma_{ij}]_{j}$, where $j\in\mathcal{J}_{\rm CPU}(\alpha_i)$, $\sigma_{ij}\geq{0}$ and $\sum_{j}\sigma_{ij}=1$. To provide an example, consider the case of 3 available CPU cores, i.e., $\mathcal{J}_{\rm CPU}(\alpha_i)=\{1,2,3\}$. In this case, a vector of the form 
$\sigma_{i} = (0.2,0.5,0.3)$ is a strategy vector, such that $20\%$ corresponds to the probability of assigning itself to CPU core $1$, $50\%$ to CPU core $2$ and $30\%$ to CPU core $3$. Briefly, the CPU core selection will be denoted by $\beta_i \in \mathcal{J}_{\rm CPU}(\alpha_i).$ Note that if $\sigma_i$ is a unit vector, say $e_j$, then agent $i$ selects its $j$th action with probability one. 

In particular, the steps executed in each iteration of the perturbed learning automata are depicted in Table~\ref{Tb:LearningAutomataForPinning}.
\begin{table}[t!]
\caption{Perturbed Learning Automata for CPU-core Pinning}
\boxed{
\begin{minipage}{0.48\textwidth}
At fixed time instances denoted by $k=1,2,...$, the following steps are executed recursively for each thread $i$ in parallel.

(1) \textbf{Performance measurement.} For the currently selected CPU-core $\beta_i(k)$ thread $i$ retrieves its current performance measurement, $\tilde{u}_i(k)$.

(2) \textbf{Strategy update.} Given that $\alpha_i$ is the current NUMA-node assignment of thread $i$, and $|\mathcal{J}_{\rm CPU}(\alpha_i)|$ is the number of the available CPU cores, the strategy of thread $i$ with respect to its CPU-core pinning is defined as:
\begin{equation}	\label{eq:SelectionRule}
\sigma_i(k) = (1-\lambda)x_i(k) - \frac{\lambda}{|\mathcal{J}_{\rm CPU}(\alpha_i)|}
\end{equation}
where $\lambda>0$ corresponds to a perturbation term (or \emph{mutation}) and $x_i(k)$ corresponds to the \emph{nominal strategy} of agent $i$. The nominal strategy is updated according to the following update recursion:
\begin{equation}	\label{eq:StrategyUpdate}
x_i(k+1) = x_i(k) + \epsilon\cdot \tilde{u}_i(k) \cdot [e_{\beta_i(k)}-x_i(k)]
\end{equation}
for some constant step-size $\epsilon>0$.

(3) \textbf{Action update.} The action of each thread $i$ is updated as follows:
$$\beta_i(k+1) = \RAND{\sigma_i}{\mathcal{J}_{\rm CPU}(\alpha_i)}.$$ 
\end{minipage}
}
\label{Tb:LearningAutomataForPinning}
\end{table}
According to this recursion, if currently thread $i$ selected CPU core $\beta_i(k)$, and measured performance $\beta_i(k)$, then its strategy is going to increase in the direction of the selected action and proportionally to the observed performance. Informally, the dynamics reinforce repeated selection and reinforcement is always proportional to the received reward.  



%
%
%
\section{Convergence Analysis}    \label{sec:ConvergenceAnalysis}

The problem of optimally allocating threads into CPU cores can be formulated as a \emph{load-balancing game}. Such formulation can help us provide immediate answer with respect to whether optimal allocations exist as well as the characteristics of these allocations. The notion of \emph{weak-acyclicity} \cite{fabrikant_structure_2013} in \emph{strategic-form games} can help us provide an answer to these questions.

In the context of load-balancing games, we are given a set of \emph{tasks} (or computing \emph{threads}) that need to be executed in a multi-core computing system (comprising multiple CPU cores). An objective may correspond to the minimization of the \emph{makespan}, that is the maximum \emph{load} over all the available CPU cores. In this case, the computing \emph{load} of a CPU core corresponds to the total computing bandwidth requested by all threads assigned to this core, that is the frequency with which the CPU core is reserved by all threads.

More formally, there exist $m$ CPU cores with \emph{speeds} $s_1, s_2, ..., s_m$ and $n$ threads with \emph{weights} $w_1, w_2, ..., w_m$, where the weight of a thread $i$ characterizes its operation/service level (e.g., the computing bandwidth requested). The speed $s_j$ of CPU core $j$ will be defined as the maximum number of instructions per sec (IPS) that can be executed by the CPU core. Moreover, the weight $w_i$ of a thread $i$ will be measured by the number of instructions per second that this thread will require within a unit of available bandwidth.

The speed $s_j$ of machine $j$ may not necessarily be known in advance (usually average over many different types of threads). Also, the weight $w_i$ may also not be available, while it may change throughout the execution time of a thread. For now, let us assume that these quantities are constant, but not necessarily known. As we will see, the explicit knowledge of these quantities will not be necessary.

\begin{figure}[h!]
\centering
\includegraphics[scale=1]{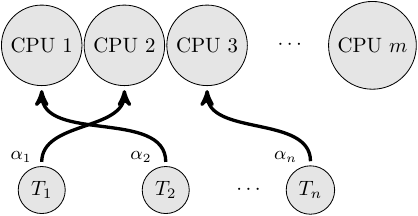}
\caption{A sketch of a load-balancing allocation problem in the context of a multi-core computing system. Each running thread independently pins itself to a single CPU core. Multiple threads may run on the same CPU core.}
\label{fig:LoadBalancing}
\end{figure}

We can analyze the problem of allocating threads into CPU cores within the context of \emph{strategic-form games}. In strategic-form games, there exists a set of players/agents $\cI\df\{1,...,n\}$, which in this case     to be the set of threads requesting resources, and $\mathcal{J}_{\rm CPU}\df\{1,...,m\}$ to be the set of machines or CPU cores available. In this setting, each thread may be thought of as an independent player that can decide independently with respect to which one of the available cores to run on. In this context, $\beta_i\in\mathcal{J}$ corresponds to the action of thread $i$, which may be any one of the available cores $\mathcal{J}_{\rm CPU}$, and $\beta\df(\beta_1,...,\beta_n)$ corresponds to the action profile over all threads (or \emph{assignment}). 

This definition of actions naturally fit to the setup of Perturbed Learning Automata for CPU-core pinning of Section~\ref{sec:PerturbedLearningAutomata}, where each thread $i$ regularly updates its selection $\beta_i$ so that threads  gradually learn the optimal allocation. Can threads, however, learn to play an optimal allocation? In order to answer this question, we need to have a closer look on the structure and properties of their interaction. Such investigation can be performed in the context of strategic-form games and it will be described in the following section.

\subsection{Weak-acyclicity and optimal CPU-core pinning}    \label{sec:WeakAcyclicity}

As it is the case in standard operating systems, each thread may run in either one of the available CPU cores under no constraints, e.g., all threads may run on the same core. However, the number of threads running on the same CPU core influences the speed with which these threads will be executed (a high number of threads on the same CPU core will lead to a low processing speed for these threads and vice versa). In particular, the \emph{load} of a CPU core $j\in\mathcal{J}$ under assignment $\beta$ will be defined as 
\begin{equation}    \label{eq:LoadBalancingLoad}
\ell_j(\beta) \df \frac{\sum_{\{k\in\mathcal{I}:\beta_k = j\}}w_k}{s_{j}} > 0.
\end{equation}
We will also denote the maximum load under profile $\beta$ as $L(\beta)\df\max_{j\in\mathcal{J}_{\rm CPU}}\ell_j(\beta_j)$. In other words, $L(\beta)$ corresponds to the \emph{makespan}, cf., \cite[Chapter~20]{nisan_selfish_2007}.

Although the speed $s_j$ of CPU core $j$ and the weight $w_i$ of thread $i$ may not be known in advance, the actual running speed of a thread on a given core can be measured in real-time quite accurately (that is the total number of completed instructions per sec which may include computational or memory related instructions). 

We define the utility of thread $i$ as the number of instructions completed per sec on core $j$, which can be expressed as follows:
\begin{equation}	\label{eq:LoadBalancingUtility}
u_i(\beta_i=j,\beta_{-i}) \df \frac{w_i}{\sum_{\{k\in\mathcal{I}:\beta_k=j\}}w_k}s_j  = \frac{w_i}{\ell_j(\beta)},
\end{equation}
where we have assumed that the operating system allocates fairly the available bandwidth in CPU core $j$ over all threads and proportionally to their weights. It is important to note that $w_i$ and $\ell_j(\beta)$ may not be known or easily measured. However, the utility $u_i$ can directly be measured on regular time intervals and per thread. Thus, it can directly be integrated into the implementation of the algorithms in Tables~\ref{Tb:AspirationLearningForNUMAPlacement}--\ref{Tb:LearningAutomataForPinning}. This design is motivated by the measurement-based optimization approach for resource allocation introduced in \cite{chasparis_measurement-based_2019}. It also introduces a slightly different design than the classical treatment of load-balancing games (see, e.g., \cite{nisan_selfish_2007}), where the cost function of a thread is defined as the load of the core. 

The strategic-form game, characterized by the tuple $\langle\mathcal{I},\mathcal{A},\{u_i\}_i\rangle$ will be referred to as a \emph{load-balancing game}. We are specifically interested in allocations that correspond to \emph{(pure) Nash equilibria}, that is allocations $\beta^*$ at which no thread would have the incentive to switch to a different CPU core. In particular, an allocation $\beta^*$ is a \emph{Nash equilibrium} if $u_i(\beta_i',\beta_{-i}^*) \leq u_i(\beta_i^*,\beta_{-i}^*)$ for all $\beta_i'\neq\beta_i^*$. 

Let us denote the set of Nash-equilibrium allocations by $\mathcal{B}_{\rm NE}$. Moreover, let us define the set $\mathcal{B}^*$ of optimal allocations as 
\begin{equation} \label{eq:OptimalAllocations}
\mathcal{B}^* \df \left\{\forall\beta\in\mathcal{B}: L(\beta^*) \leq L(\beta) \right\}.
\end{equation}
In other words, the set of optimal assignments minimizes the makespan. Let also denote $L^*$, the minimum makespan that can be achieved at the optimal assignments.
\begin{proposition}[Existence of Nash equilibria]
Consider the load-balancing game characterized by the tuple $\langle\mathcal{I},\mathcal{A},\{u_i\}_i\rangle$ with a utility function defined by (\ref{eq:LoadBalancingUtility}). Then, the set of pure Nash equilibria is non-empty, i.e., $\mathcal{B}_{\rm NE}\neq\varnothing$.
\end{proposition}
\begin{proof}
Let us consider any allocation profile $\beta$ which is not a pure Nash equilibrium. In other words, there exists a thread $i$ and two available CPU cores $j$ and $l$, such that, switching from core $j$ to core $l$ \emph{strictly} increases the utility of thread $i$ (i.e., its processing speed). In particular, given that:
\begin{align}
u_i(\beta_i=j,\beta_{-i}) - u_i(\beta_i'=l,\beta_{-i}) & = w_i\frac{\ell_l(\beta') - \ell_j(\beta)}{\ell_l(\beta')\ell_j(\beta)}
\end{align} 
we conclude that, if $u_i(\beta')>u_i(\beta)$ (i.e., $\beta'$ is a better reply to $\beta$) then $\ell_j(\beta) > \ell_l(\beta')$. In other words, if thread $i$ strictly improves its speed by switching from core $j$ to core $l$, it implies that the load of core $j$ (when $i$ runs on core $j$) is strictly larger than the load of core $l$ (when $i$ runs on core $l$). Thus, we conclude that $L(\beta')\leq L(\beta)$, i.e., under any better reply, the makespan reduces or remains the same. Furthermore, the number of threads that have a load which is equal or higher than $\ell_j(\beta)$ has now been strictly decreased. We conclude that this process may only terminate at a state than no thread can improve its speed any further, i.e., at a Nash equilibrium.
\end{proof}

The importance of this proposition lies on the fact that there exists a set of Nash equilibria at which all threads perform well at least locally. Note that the set of Nash equilibria may not necessarily coincide with the set of optimal allocations $\mathcal{B}^*$. In fact, the set of optimal allocations may or may not be part of the set of Nash equilibria. However, certain guarantees can be established with respect to the utility achieved at the worst Nash equilibrium as compared to the utility received at an optimal allocation. The following proposition provides a lower bound on the performance of any Nash equilibrium as compared to the performance of an optimal assignment. We only investigate the case of identical CPU cores, since this condition is satisfied by our experimental setup.

\begin{proposition}[Performance of Nash equilibria]     \label{Pr:PerformanceNashEquilibria}
For the case of identical CPU cores and for any pure Nash equilibrium assignment $\beta\in\mathcal{B}_{\rm NE}$, the makespan satisfies
\begin{equation}    \label{eq:UpperBoundMakespan}
    L(\beta) \leq \frac{2\magn{\mathcal{J}_{\rm CPU}}}{\magn{\mathcal{J}_{\rm CPU}}+1} \cdot L^*
\end{equation}
where $\magn{\mathcal{J}_{\rm CPU}}$ denotes the number of available CPU cores. Furthermore, the utility of any thread $i\in\cI$ at any pure Nash equilibrium assignment $\beta\in\mathcal{B}_{\rm NE}$ satisfies
\begin{equation}    \label{eq:UpperBoundUtility}
    u_i(\beta) \geq \frac{\left(\magn{\mathcal{J}_{\rm CPU}}+1\right)}{2\magn{\mathcal{J}_{\rm CPU}}} \cdot \frac{w_i}{L^*}.
\end{equation}
\end{proposition}
\begin{proof}
The proof of the first statement (\ref{eq:UpperBoundMakespan}) follows the exact same reasoning with Theorem~20.5 in \cite{nisan_selfish_2007}. The proof of the second statement (\ref{eq:UpperBoundUtility}) follows directly from the definition of the utility (\ref{eq:LoadBalancingUtility}) and the first statement (\ref{eq:UpperBoundMakespan}). In particular, let us consider any thread $i$ with weight $w_i$. Its speed will satisfy:
\begin{equation*}
    u_i \geq \frac{w_i}{L(\beta)} \geq \frac{\left(\magn{\mathcal{J}_{\rm CPU}}+1\right)}{2\magn{\mathcal{J}_{\rm CPU}}} \cdot \frac{w_i}{L^*}.
\end{equation*}
which concludes the proof.
\end{proof}

The above proposition provides a lower-bound in the utility that can be achieved at a Nash equilibrium assignment. In particular, note that the ratio $u_i^*\df\nicefrac{w_i}{L^*}$ corresponds to the least maximum speed that a thread can achieve under an optimal assignment. Thus, in a 10 CPU-core architecture, condition (\ref{eq:UpperBoundUtility}) implies that $u_i(\beta)\geq \nicefrac{11}{20}u_i^*$. Such lower bound is a bit conservative, however it provides a significant guarantee. 

From Equation~(\ref{eq:UpperBoundUtility}), we may also conclude that:
\begin{equation*}
    \frac{1}{\magn{\mathcal{J}_{\rm CPU}}}\sum_{i\in\cI}u_i \geq \frac{\left(\magn{\mathcal{J}_{\rm CPU}}+1\right)}{2\magn{\mathcal{J}_{\rm CPU}}} \cdot \left(\frac{1}{\magn{\mathcal{J}_{\rm CPU}}} \sum_{i\in\mathcal{I}} \frac{w_i}{L^*}\right),
\end{equation*}
which also establishes a similar lower bound with respect to our original (desirable) objective of maximizing the average speed over all threads. 

We conclude that if threads settle on a Nash equilibrium assignment, then there is a certain guarantee with respect to their average running speed.

\subsection{Convergence analysis of CPU-core pinning}      \label{sec:ConvergenceCPU}

The previous section discussed existence and properties of assignments that are Nash equilibria of the load balancing game of the CPU-core assignment problem. Given the properties of Proposition~\ref{Pr:PerformanceNashEquilibria}, Nash-equilibrium assignments should be desirable, since they provide certain guarantees with respect to the overall performance. However, \emph{can the dynamics presented in Section~\ref{sec:DynamicScheduler} of Tables~\ref{Tb:AspirationLearningForNUMAPlacement}--\ref{Tb:LearningAutomataForPinning} guarantee convergence to the set of Nash-equilibrium assignments?} This is the question we try to answer in this section.

First, we will investigate the convergence properties of the dynamics of Table~\ref{Tb:LearningAutomataForPinning} under the condition of a single NUMA-node availability. In other words, threads do not have the opportunity to migrate, and they can only increase their utility by improving their pinning assignment to the available CPU cores. The following proposition provides strong guarantees with respect to the convergence of the dynamics for CPU-core pinning of Table~\ref{Tb:LearningAutomataForPinning}.

\begin{proposition}[Convergence of CPU-pinning]     \label{Pr:ConvergenceCPUPinning}
Consider the update recursion of Table~\ref{Tb:LearningAutomataForPinning}. The fraction of time that the discrete-time dynamics spends in an arbitrarily small neighborhood of the set of pure Nash equilibria goes to one as the perturbation factor $\lambda\downarrow{0}$, the step-size $\epsilon\downarrow{0}$ and the time index $k\to\infty$.
\end{proposition}
\begin{proof}
Theorem~3.1 in \cite{chasparis_stochastic_2019} has shown that as the perturbation factor $\lambda\downarrow{0}$, the induced Markov chain of the dynamics of Table~\ref{Tb:LearningAutomataForPinning} has an invariant probability measure whose support lies on the pure strategy states (i.e., states at which for all $i$, $x_i$ assigns probability one to some action). By Birkhoff's individual ergodic theorem \cite[Theorem~2.3.4]{Lerma03}, this implies that the process will spend an arbitrarily large portion of time on pure-strategy states as $\lambda\downarrow{0}$ and $k\to\infty$. Furthermore, according to \cite[Proposition~3.6]{ChasparisShamma11_DGA},  $\lambda$-perturbations of pure Nash equilibria are the unique limit points of the continuous-time approximation of the dynamics (\ref{eq:StrategyUpdate}). Thus, according to a straightforward implementation of \cite[Theorem~8.2.1]{KushnerYin03}, the fraction of time that the discrete-time dynamics (\ref{eq:StrategyUpdate}) spends in a small neighborhood of the set of pure Nash equilibria goes to one as $\epsilon\downarrow{0}$ and $k\to\infty$.
\end{proof}

\subsection{Discussion on combined NUMA and CPU placements}

The main motivation for decomposing the decision making process into NUMA-placement and CPU-pinning in Tables~\ref{Tb:AspirationLearningForNUMAPlacement}--\ref{Tb:LearningAutomataForPinning}, respectively, lies on the principle of the two time-scale dynamics. In particular, the NUMA placement algorithm of Table~\ref{Tb:AspirationLearningForNUMAPlacement} operates at a slow time-scale with a period of $T_{\rm NUMA}$, while the CPU-pinning of Table~\ref{Tb:LearningAutomataForPinning} operates at a faster time-scale with a period $T_{\rm CPU}\ll T_{\rm NUMA}$. The goal is to allow the dynamics of CPU-pinning to first approach a Nash-equilibrium assignment (given the convergence guarantees of Proposition~\ref{Pr:ConvergenceCPUPinning}), before any thread considers migrating to a different NUMA node. Such design principle also restricts frequent NUMA-node migrations, since they may be rather costly (taking into account possible implications to memory access).

When we select $T_{\rm NUMA}/T_{\rm CPU}$ to be sufficiently large, then the CPU-core pinning dynamics have already settled in the set of pure Nash equilibria (according to Proposition~\ref{Pr:ConvergenceCPUPinning}) before revising the migration of threads to different nodes. There are two possibilities that a thread decides to migrate. Under the first condition (4a) of Table~\ref{eq:AspirationLevelUpdateRule}, thread $i$ is unsatisfied under the current assignment, and randomly selects among alternative NUMA nodes where currently threads perform better on average. By appropriately selecting sufficiently small $\gamma\in(0,1)$ in the implementation of the better-reply condition (\ref{eq:NUMABetterReplyCondition}), a migration to a new NUMA node will only result in an increased processing speed for a thread. This is also guaranteed by the fact that only one thread is allowed to migrate at a given time. Under the second condition (4b) of Table~\ref{eq:AspirationLevelUpdateRule}, there always exists a small probability $\zeta>0$ that a (neither satisfied nor unsatisfied) thread is selected to migrate at random and given that there are alternative nodes that can offer a better performance. Thus, under either condition, and for sufficiently large $T_{\rm NUMA}/T_{\rm CPU}$, we should expect that threads may only increase their performance by migrating.

\section{Experiments}		\label{sec:Experiments}

In this section, we present an experimental study of the proposed framework. Experiments were conducted on \texttt{20$\times$Intel\copyright Xeon\copyright CPU E5-2650 v3 \@ 2.30 GHz} running Linux Kernel 64bit 3.13.0-43-generic. The cores are divided into two NUMA nodes (Node 1: 0-9 CPU cores, Node 2: 10-19 CPU cores). 

In all experiments, the utility of each thread is defined as the \emph{total instructions completed per second} which incorporates both the computational and memory-access instructions. This is a multi-objective criterion and it is expected that the larger the number of instructions completed, the larger the processing speed of a thread. We compared the overall performance of the application (in terms of processing speed of threads and completion time of an application) with that of the Linux \OS\ scheduler. We considered a number of parallel applications under different levels of resource availability (i.e., number of CPU cores available for the applications) and background-load settings (i.e., number of threads of other applications running on the available cores at the same time).

\subsection{Benchmark applications}

In particular, we have considered the following benchmark applications:
\begin{itemize}
\item \emph{Swaptions} (SWA), that uses the Heath-Jarrow-Morton (HJM) framework to price a portfolio of swaptions. The HJM framework describes how interest rates evolve for risk management and asset liability management~\cite{HJM}. The application employs Monte-Carlo simulation to compute the prices. It is regular in terms of task sizes, with a low degree of communication between different threads. It was taken from the \emph{Parsec} benchmark suite. 
\item \emph{Blackscholes} (BLA), that calculates, using differential equations, how the value of an option changes as the price of the underlying asset changes; parallel implementation calculates values for a number of options at the same time, assigning a thread to each option (or a group of options). If the options are equally divided between threads, this results in a regular (in terms of task sizes) parallel application. In practice, similar calculations are used by financial houses to price 10-100 thousands of options. This is \emph{computationally intensive} application as depicted in Table~\ref{tb:InstensityACO}. It was taken from the \emph{Parsec} benchmark suite. 
\item \emph{Ant Colony Optimization (ACO)}
\cite{Dorigo-aco-book} is a metaheuristic used for solving NP-hard combinatorial optimization problems.  In this paper, we apply ACO to the Single Machine Total Weighted Tardiness Problem (SMTWTP). Briefly, this is a scheduling problem of jobs that are characterized by varying processing times, deadlines and weights. The objective is to find the schedule that minimizes the total tardiness. A detailed description of this use case is provided in \cite{chasparis_euro-par_2017}.
This is \emph{computationally intensive} application as depicted in Table~\ref{tb:InstensityACO}.

\begin{table}[t!]
\centering
\caption{Computational/Memory Intensity of Case Studies ({\rm TOT\_INS} = total instructions, {\rm LST\_INS} = load/store instructions, {\rm TLB\_DM} = Data translations) }
\begin{tabular}{|c|c|c|c|c|}
\hline
Index & BLA & SWA & ACO & CSO \\\hline
{\rm TOT\_INS} / {\rm LST\_INS} & $\mathcal{O}(10^{+7})$ & $\mathcal{O}(10^{+6})$ & $\mathcal{O}(10^{+5})$ & $\mathcal{O}(10^{+2})$\\ \hline
{\rm TLB\_DM} / {\rm LST\_INS} & $\mathcal{O}(10^{-7})$ & $\mathcal{O}(10^{-6})$ & $\mathcal{O}(10^{-5})$ & $\mathcal{O}(10^{-2})$\\ \hline
\end{tabular}
\label{tb:InstensityACO}
\end{table}

\item \emph{Stochastic-Local-Search for Cutting-Stock Industrial Optimization (CSO)} that optimizes classical bin-packing and cutting-stock optimization problems using an evolutionary stochastic-local-search (SLS) algorithm. The use case and the type of parallelization (which is based on the Fast-Flow parallelization library \cite{aldinucci_pool_2016}) has been described in detail in \cite{DBLP:journals/corr/ChasparisRH17}. In particular, we used the Scholl 1--3 datasets for classical bin packing problems provided in \cite{BinPackingLibrary}. According to the implemented SLS algorithm, an initial number of candidate solutions (pool) of a bin-packing/cutting-stock problem, are  processed continuously through a series of heuristic based operations/modifications (optimization cycle). In each such cycle, multiple threads are assigned a portion of the candidate solutions. Since the application usually runs for a fixed time, the total number of candidate solutions processed in all optimization cycles completed constitutes an indication of the average processing speed. This is a \emph{memory intensive} application as depicted in Table~\ref{tb:InstensityACO}, while the computation bandwidth requested varies significantly with time. 

\end{itemize}


\subsection{Experimental setup} 

The period of the CPU pinning is fixed to $T_{\rm CPU}=0.05$ sec, which is also the interval in which the \RM\ collects measurements of the \emph{total instructions completed per sec} (using the PAPI library \cite{Mucci99}) for each one of the threads separately. In other words, the \emph{utility} $u_i$ of thread $i$ corresponds to the total instructions completed per sec for thread $i$.

Pinning of threads to CPU cores is achieved through the \texttt{sched.h} library. 
In all experiments, the \RM\ is executed by the master thread of an application, which is always running in a fixed CPU core (usually the first available CPU core of the first NUMA node). 

In Table~\ref{Tb:ACOExperiments}, we provide an overview of the conducted experiments. We classify the experiments with respect to the resource availability and the CPU availability. We classify the resource availability as \emph{small} (around 4 application threads per CPU core), \emph{medium} (2 threads per CPU core) and \emph{high} (1 thread per CPU core). We classify the CPU availability as \emph{uniform}, when no background applications are running and therefore all CPU cores are fully available to the tested application, \emph{non-uniform} where 8 threads of a background application are running on the first 4 CPU cores of the machine for the whole duration of the running of the tested application and \emph{time-varying}, where initially the availability varies continuously with time in the first 4 CPU cores of the machine.

Our goal is to investigate the performance of the scheduler under different set of available resources, and how the dynamic scheduler adapts to background load.  

\begin{table}[t!]
\centering
\caption{Algorithm settings}
\begin{tabular}{|c|c|}
\hline
Parameter & Value \\\hline\hline
$\epsilon$ & $0.01/10^{8}$ \\\hline
$\lambda$ & $0.02$ \\ \hline
$T_{\rm CPU}$ & $0.05$ sec \\ \hline\hline
$\nu$ & $0.01$ \\ \hline
$\zeta$ & $0.02$ \\ \hline
$\gamma$ & $0.9$ \\ \hline
$\eta$ & $0.8$ \\\hline
$T_{\rm NUMA}$ & $2$ sec \\\hline
\end{tabular}
\label{tb:AlgorithmSettings}
\end{table}

\begin{table}[tbh!]
\caption{Classification of the experiments.}
\centering
\begin{tabular}{c|c|c}
\centering
\textbf{Exp.}	& 	\textbf{Resource availability} & \textbf{CPU availability}  		\\\hline\hline
A.1 			&  Small 	& Uniform \\\hline
A.2			&   Small	 & Non-uniform \\\hline
A.3 			&   Small	& Time-varying \\\hline\hline
B.1			&   Medium  & Uniform \\
\hline
B.2			&   Medium  & Non-uniform\\
\hline
B.3			&   Medium & Time-varying \\\hline\hline
C.1			&   Large  & Uniform \\
\hline
C.2			&   Large  & Non-uniform \\
\hline
C.3			&   Large  & Time-varying \\\hline\hline
\end{tabular}
\vspace{0.1cm}
\label{Tb:ACOExperiments}
\end{table}

\subsection{Experimental Results} \label{sec:ExperimentalResults}

\begin{table}[tbh!]
\centering
\caption{Completion times of \OS\ and \RL\ scheduling for Swaptions application. We show the mean execution time of the application, the deviation and improvement in execution time of \RL\ over \OS\ scheduling}
\label{Tb:ExperimentsSWO}
\begin{tabular}{|c||c|c||c|c||c|}
\hline
\multirow{2}{*}{\begin{minipage}{0.09\textwidth} \centering \textbf{Exp/}\\ \textbf{Resources} \end{minipage}}	& 	\multicolumn{2}{|c||}{\textbf{\OS}}  & \multicolumn{2}{|c||}{\textbf{\RL}} & \multirow{2}{*}{\begin{minipage}{0.10\textwidth} \centering \textbf{Diff. (\%)} \end{minipage}} \\
  \cline{2-5} & Mean  & Dev & Mean & Dev &  \\ \hline \hline
  SWA (A.1) & \textbf{225.58} & 1.28 &  \textbf{225.27} & 1.41 &  $\mathbf{+0.13}$ \\ \hline
  SWA (A.2) & \textbf{385.75} & 17.00 &  \textbf{344.53} & 3.38 &  $\mathbf{+10.69}$ \\ \hline
  SWA (A.3) & \textbf{337.46} & 14.62 &  \textbf{311.17} & 2.98 &  $\mathbf{+7.79}$ \\ \hline \hline
  \hline
  SWA (B.1) & \textbf{163.40} & 0.56  &  \textbf{158.10} & 2.20   & $\mathbf{+3.25}$ \\ \hline
  SWA (B.2) & \textbf{289.31} & 5.93 &   \textbf{285.68} & 5.28 &  $\mathbf{+1.26}$ \\ \hline
  SWA (B.3) & \textbf{240.81} & 5.22 &   \textbf{238.05} & 4.89 & $\mathbf{+1.15}$ \\ \hline \hline
 \hline
  SWA (C.1) & \textbf{122.54} & 0.79 &  \textbf{129.85} & 3.25 &  $\mathbf{-5.96}$ \\ \hline
  SWA (C.2) & \textbf{206.68} & 1.94 &  \textbf{202.85} & 2.83 &  $\mathbf{+1.85}$ \\ \hline
  SWA (C.3) & \textbf{164.11} & 1.49 & \textbf{161.54} & 3.19 &  $\mathbf{+1.57}$ \\ \hline
\end{tabular}
\end{table}

\begin{table*}[tbh!]
\centering
\caption{Completion times (CT) and average processing speed (Avg. Spd) of \OS\ and \RL\ scheduling for ACO application. We show the mean execution time of the application, mead deviation (in seconds) and average processing speed per thread (in $10^8$ instructions per second).}
\label{Tb:ExperimentsACO}
\begin{tabular}{|c||c|c|c||c|c|c||c|c|}
\hline
\multirow{3}{*}{\begin{minipage}{0.11\textwidth} \centering \textbf{Exp/} \\ \textbf{Time(s)} \end{minipage}}	& 	\multicolumn{3}{|c||}{\textbf{\OS}}  & \multicolumn{3}{|c||}{\textbf{\RL}} & \multirow{2}{*}{\begin{minipage}{0.11\textwidth} \centering \textbf{Diff. CT} \\ \textbf{(\%)} \end{minipage}} & \multirow{2}{*}{\begin{minipage}{0.11\textwidth} \centering \textbf{Diff. Avg. Spd (\%)} \end{minipage}}\\
  \cline{2-7} & Mean CT & Dev CT & Avg. Spd. & Mean CT & Dev CT & Avg. Spd & & \\ \hline
  ACO (A.1) & \textbf{1065.05} & 7.68 & 13.37 & \textbf{1075.48} & 6.45 & 14.87 & $\mathbf{-0.9}$ & $\mathbf{+11.21}$ \\ \hline
  ACO (A.2) & \textbf{1752.46} & 14.00 & 8.54 & \textbf{1455.92} & 22.8 & 9.82 & $\mathbf{+16.92}$ & $\mathbf{+14.98}$ \\ \hline
  ACO (A.3) & \textbf{1459.18} & 9.42 & 10.29 & \textbf{1402.00} & 4.06 & 10.41 & $\mathbf{+3.91}$ & $\mathbf{+1.17}$ \\ \hline \hline
  ACO (B.1) & \textbf{673.09} & 5.69 & 21.26 & \textbf{699.16} & 10.24 & 22.16 & $\mathbf{-3.87}$ & $\mathbf{+4.23}$ \\ \hline
  ACO (B.2) & \textbf{1106.36} & 16.71 & 12.73 & \textbf{1041.33} & 16.71 & 14.94 & $\mathbf{+5.87}$ & $\mathbf{+17.36}$ \\ \hline
  ACO (B.3) & \textbf{1066.18} & 0.88 & 13.20 & \textbf{1019.11} & 8.39 & 14.58 & $\mathbf{+4.41}$ & $\mathbf{+10.45}$ \\ \hline \hline
  ACO (C.1) & \textbf{455.87} & 5.08 & 31.90 & \textbf{496.26} & 5.08 & 33.46 & $\mathbf{-8.85}$ & $\mathbf{+4.89}$\\ \hline
  ACO (C.2) & \textbf{659.78} & 27.45 & 21.57 & \textbf{688.80} & 18.66 & 24.15 & $\mathbf{-4.39}$ & $\mathbf{-12.02}$ \\ \hline
  ACO (C.3) & \textbf{659.35} & 3.62 & 21.82 & \textbf{676.03} & 7.72 & 23.72 & $\mathbf{-2.52}$ & $\mathbf{+8.70}$ \\ \hline\hline
 \multicolumn{6}{c}{} & \textbf{Average} & $\mathbf{+1.17}$ & $\mathbf{+6.77}$ \\ 
 \cline{8-9}
\end{tabular}
\end{table*}

\begin{table}[tbh!]
\centering
\caption{Completion times of \OS\ and \RL\ scheduling for Blackscholes (BLA) application}
\label{Tb:ExperimentsBLA}
\begin{tabular}{|c||c|c||c|c||c|}
\hline
\multirow{2}{*}{\begin{minipage}{0.09\textwidth} \centering \textbf{Exp/}\\ \textbf{Resources} \end{minipage}}	& 	\multicolumn{2}{|c||}{\textbf{\OS}}  & \multicolumn{2}{|c||}{\textbf{\RL}} & \multirow{2}{*}{\begin{minipage}{0.10\textwidth} \centering \textbf{Diff. (\%)} \end{minipage}} \\
  \cline{2-5} & Mean  & Dev & Mean & Dev &  \\ \hline \hline
  BLA (A.1) & \textbf{193.20} & 1.89 &  \textbf{190.43} & 0.62 &    $\mathbf{+1.09}$ \\ \hline
  BLA (A.2) & \textbf{322.32} & 4.98 &  \textbf{314.73} & 8.40  & $\mathbf{+2.36}$ \\ \hline
  BLA (A.3) & \textbf{285.76} & 4.17 &  \textbf{274.17} & 7.30 &  $\mathbf{+4.05}$ \\ \hline \hline
  BLA (B.1) & \textbf{129.98} & 1.09  & \textbf{129.88} & 1.31 &  $\mathbf{+0.08}$ \\ \hline
  BLA (B.2) & \textbf{236.62} & 4.18 &  \textbf{245.09} & 2.64 & $\mathbf{-3.58}$ \\ \hline
  BLA (B.3) & \textbf{192.45} & 5.16 &  \textbf{200.15} & 4.46 &  $\mathbf{-4.00}$ \\ \hline \hline
  BLA (C.1) & \textbf{98.97} & 1.11 &  \textbf{107.77} & 1.25 &  $\mathbf{-8.89}$ \\ \hline
  BLA (C.2) & \textbf{166.50} & 1.46 &  \textbf{172.65} & 3.00 &  $\mathbf{-3.69}$ \\ \hline
  BLA (C.3) & \textbf{130.24} & 2.13 &  \textbf{135.42} & 3.87 &  $\mathbf{-3.98}$ \\ \hline
\end{tabular}
\end{table}

\begin{table*}[tbh!]
\centering
\caption{Candidate solutions processed (CSP) and average processing speed (Avg. Spd) under \OS\ and \RL\ scheduling for CSO application within 10min simulation time. We show the Mean solutions processes, the deviation, and average processing speed per thread (in $10^8$ instructions per second).}
\label{Tb:ExperimentsCSO}
\begin{tabular}{|c||c|c|c||c|c|c||c|c|}
\hline
\multirow{2}{*}{\begin{minipage}{0.09\textwidth} \centering \textbf{Exp/}\\ \textbf{Resources} \end{minipage}}	& \multicolumn{3}{|c||}{\textbf{\OS}}  & \multicolumn{3}{|c||}{\textbf{\RL}} & \multirow{2}{*}{\begin{minipage}{0.12\textwidth} \centering \textbf{Diff. CSP \\ (\%)} \end{minipage}} & \multirow{2}{*}{\begin{minipage}{0.12\textwidth} \centering \textbf{Diff. Avg. Spd (\%)}\end{minipage}} \\
  \cline{2-7} & Mean CSP & Dev CSP  & Avg. Spd & Mean CSP & Dev CSP  & Avg. Spd & & \\ \hline \hline
  CSO (A.1) & \textbf{968.80} & 20.57 & \textbf{11.61} & \textbf{965.60} & 24.10 & \textbf{9.49} & $\mathbf{-0.33}$ & $\mathbf{-18.26}$ \\ \hline
  CSO (A.2) & \textbf{398.40} & 12.07 & \textbf{5.32} & \textbf{461.60} & 10.29 & \textbf{7.09} & $\mathbf{+15.86}$ & $\mathbf{+33.27}$ \\ \hline
  CSO (A.3) & \textbf{572.80} & 9.39 & 6.93 & \textbf{577.00} & 0.00 & 7.10 & $\mathbf{+0.73}$ & $\mathbf{+2.45}$ \\ \hline\hline
  CSO (B.1) & \textbf{955.80} & 57.32 & 11.78 & \textbf{960.80} & 48.54 & 12.75 & $\mathbf{+0.52}$ & $\mathbf{+8.23}$ \\ \hline
  CSO (B.2) & \textbf{812.40} & 129.60 & \textbf{10.88} & \textbf{644.40} & 33.44 & \textbf{8.24} & $\mathbf{-20.68}$ & $\mathbf{-24.26}$ \\ \hline
  CSO (B.3) & \textbf{955.20} & 89.04 & \textbf{11.07} & \textbf{769.40} & 56.00 & \textbf{9.20} & $\mathbf{-19.45}$ & $\mathbf{-16.90}$ \\ \hline\hline
  CSO (C.1) & \textbf{925.80} & 35.58 & \textbf{10.74} & \textbf{983.20} & 41.75 & \textbf{12.62} & $\mathbf{+6.20}$ & $\mathbf{+17.50}$ \\ \hline
  CSO (C.2) & \textbf{614.80} & 14.67 & \textbf{8.74} & \textbf{616.00} & 8.94 & \textbf{8.76} & $\mathbf{+0.20}$ & $\mathbf{+0.23}$ \\ \hline
  CSO (C.3) & \textbf{746.50} & 37.47 & \textbf{8.86} & \textbf{876.60} & 72.05 & \textbf{9.05} & $\mathbf{+17.43}$ & $\mathbf{+2.14}$ \\ \hline\hline
 \multicolumn{6}{c}{} & \textbf{Average} & $\mathbf{+0.06}$ & $\mathbf{+0.48}$ \\ 
 \cline{8-9}
\end{tabular}
\end{table*}

Tables~\ref{Tb:ExperimentsSWO}--\ref{Tb:ExperimentsCSO} show the execution times of the four chosen applications under \OS\ and \RL\ scheduler and under the experimental scenarios of Table~\ref{Tb:ACOExperiments}. Below, we analyze each application separately.

\paragraph{SWA} We observe that the \RL\ scheduler exhibits better behavior than the \OS\ under small and medium availability of resources (i.e., categories A and B) with or without background interference. The improvement varies between 0.13\% and 10.69\%. In case of large availability of resources (i.e., category C), the \OS\ outperforms the \RL\ but only in the case where there is no background interference. Note also that the percentages of the deviations are significantly smaller than the corresponding performance differences (except for the A.1 case), thus we may not attribute these improvements to noise.


\paragraph{ACO} In this set of experiments, we see a similar behavior to the SWA experiments. The \RL\ outperforms the \OS\ in the case of small and medium availability of resources and in the presence of background interference (i.e., categories A.2--A.3 and B.2--B.3). The improvement may reach up to 16.92\%. In the absence of any background interference, the behavior under small availability of resources (i.e., category A.1) is about equivalent, while in the remaining categories the \OS\ outperforms the \RL\ scheduler. 

\begin{figure}[th!]
\centering
\includegraphics[scale=1]{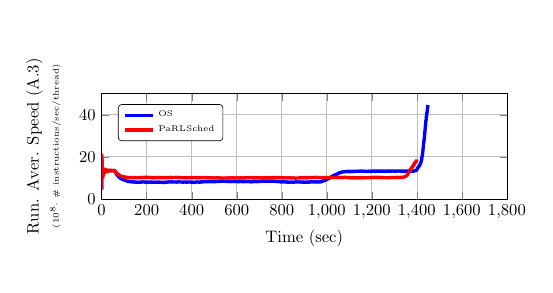}\vspace{-50pt}
\includegraphics[scale=1]{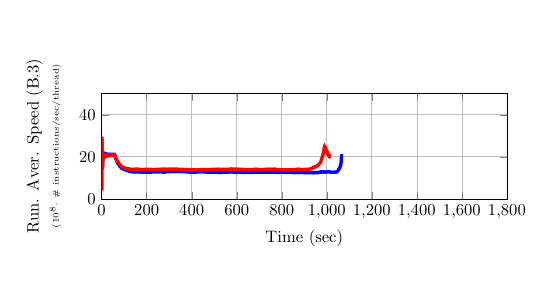}\vspace{-50pt}
\includegraphics[scale=1]{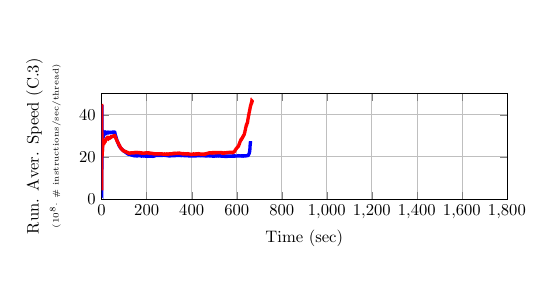}\vspace{-30pt}
\caption{Sample responses for Experiments of category ${3}$ (i.e., under time-varying CPU availability. The running average speed is measured in $( 10^8 \cdot$ $\#$ instructions/sec/thread).}
\label{fig:ExperimentSetC:SampleResponses}
\end{figure}

As a side note, we should mention that even under scenarios where the \OS\ outperformed \RL, such as scenario C.3, the average speed over all threads is not necessarily smaller, as  Figure~\ref{fig:ExperimentSetC:SampleResponses} demonstrates. In other words, the \RL\ does indeed achieve a good level of the average processing speed, which agrees with its design criterion, but apparently completion time is not only a matter of average speed. For example, a large average speed over all threads does not necessarily guarantee that all threads are running with identical speeds. Instead, there might be significant differences in the speeds between threads, which may have an impact on the overall completion time.  

\paragraph{BLA} The performance under the Blackscholes application is not deviating significantly in comparison with the conclusions of ACO and SWA applications. In fact, we observe a constantly better performance of the \RL\ in conditions of small resource availability which may reach up to 4.05\% improvement. On the other hand, the performance under large resource availability has been up to -8.89\% worse than the \OS\ performance.

\begin{figure}[th!]
\centering
\includegraphics[scale=1]{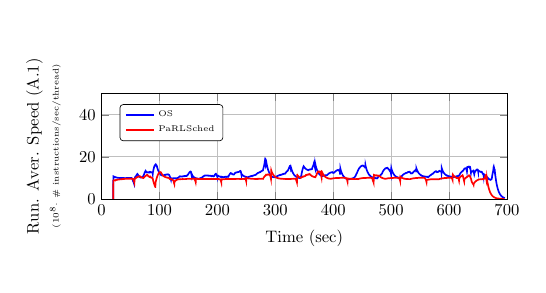}\vspace{-50pt}
\includegraphics[scale=1]{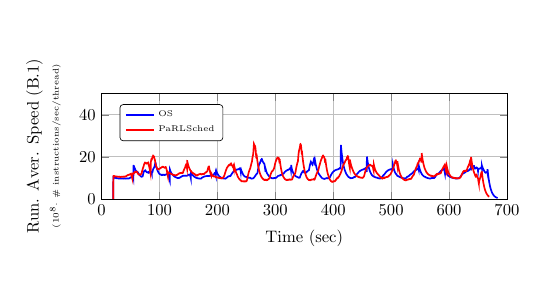}\vspace{-50pt}
\includegraphics[scale=1]{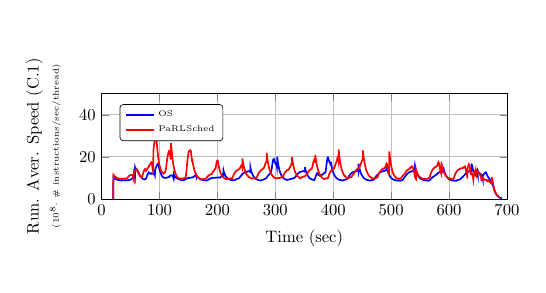}\vspace{-30pt}
\caption{Sample responses for Experiments of category ${3}$ (i.e., under time-varying CPU availability. The running average speed is measured in $( 10^8 \cdot$ $\#$ instructions/sec/thread).}
\label{fig:CSOSampleResponses1}
\end{figure}

\begin{figure}[th!]
\centering
\includegraphics[scale=1]{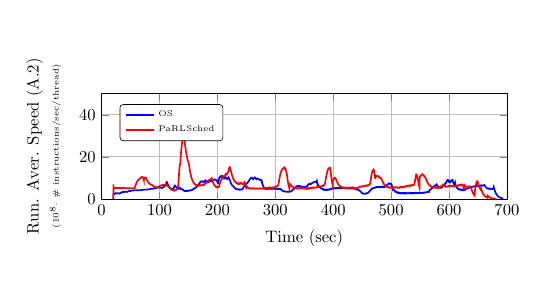}\vspace{-50pt}
\includegraphics[scale=1]{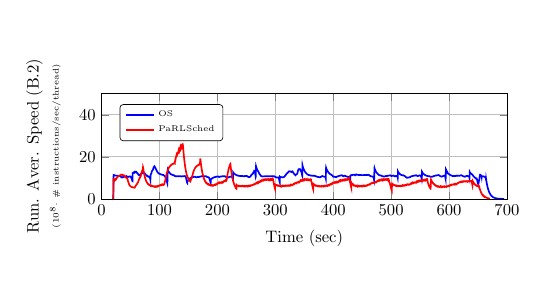}\vspace{-50pt}
\includegraphics[scale=1]{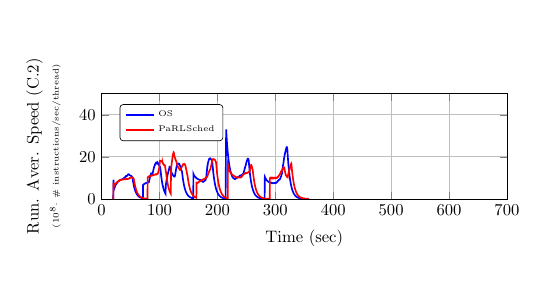}\vspace{-30pt}
\caption{Sample responses for Experiments of category ${3}$ (i.e., under time-varying CPU availability. The running average speed is measured in $( 10^8 \cdot$ $\#$ instructions/sec/thread).}
\label{fig:CSOSampleResponses2}
\end{figure}

\begin{figure}[th!]
\centering
\includegraphics[scale=1]{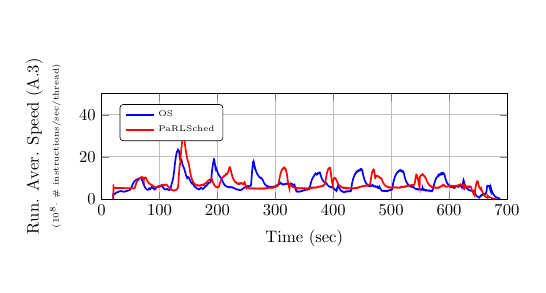}\vspace{-50pt}
\includegraphics[scale=1]{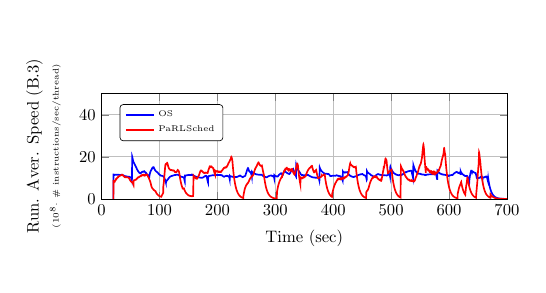}\vspace{-50pt}
\includegraphics[scale=1]{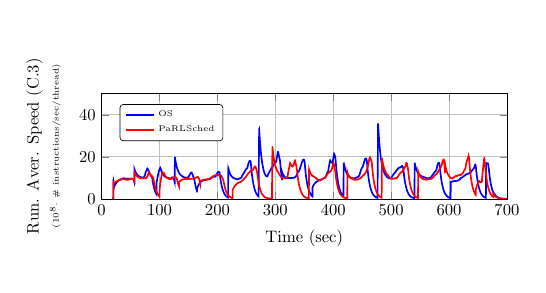}\vspace{-30pt}
\caption{Sample responses for Experiments of category ${3}$ (i.e., under time-varying CPU availability. The running average speed is measured in $( 10^8 \cdot$ $\#$ instructions/sec/thread).}
\label{fig:CSOSampleResponses3}
\end{figure}

\paragraph{CSO} The CSO application is a bit different than the ones previously considered. It is characterized by scattered memory pages as Table~\ref{tb:InstensityACO} reflects. In general, we observe significant advantage of the \RL\ scheduler under categories A and C of resource availability, and a reduced performance in the case of category B (medium availability). The rather inconclusive behavior should be attributed to the irregular memory accesses of the application and the long idle times of the threads. This large variation in the requested bandwidth is also demonstrated in Figures~\ref{fig:CSOSampleResponses1}, \ref{fig:CSOSampleResponses2}, and \ref{fig:CSOSampleResponses3} which show the response of the \RL\ scheduler under all scenarios.

\subsection{Discussion} 

In general, we observed that the \RL\ scheduler was able to achieve better performance that the OS scheduler in limited cases of limited availability of resources (Category A) and external disturbances. Under such scenarios, we expect the performance of individual threads to vary due to external influences and, therefore, it is important to make the correct remapping decisions. Also, under such scenarios, it is not possible to predict this variation in the performance solely based on the characteristics of the application itself. Finally, in the memory-intensive application (CSO), the scheduler was able to better adapt to the irregularity in the memory-access speeds between the two NUMA nodes also under large availability of resources. 

On the other hand, the \OS\ outperformed the \RL\ scheduler in most cases of large availability of resources (e.g., category C.1). This should be attributed to the fact that the Linux scheduler is utilizing internal load balancing of threads between cores, which has notable effect on the execution time when there is not significant background interference (in terms of additional running applications). In this case, performance of the individual threads depends exclusively on the distribution of threads of the application to cores, so there is no additional benefit in measuring external interference in the \RL\ scheduler. The \RL\ scheduler applies rigid pinning of threads to cores, which means that it cannot utilize any internal load balancing by the Linux scheduler. 

Given the rather diverse nature of the considered applications, the observed improvements constitute a promising indication. Note that the intention and goal of this work is not to replace the \OS\ scheduler, but instead to act on a supervisory level, and possibly under alternative multi-objective criteria. The notion of the utility function that drives the thread placement can be designed to accommodate any such multi-objective criterion, since the only assumption considered is the positivity constraint.

\section{Conclusions and future work} \label{sec:Conclusions}

We proposed a measurement- (or performance-)  based learning scheme for addressing the problem of efficient dynamic pinning of parallelized applications into many-core systems under a NUMA architecture. According to this scheme, a centralized objective is decomposed into thread-based objectives, where each thread is assigned its own utility function. Allocation decisions were organized into a hierarchical decision structure: at the first level, decisions are taken with respect to the assigned NUMA node, while at the second level, decisions are taken with respect to the assigned CPU core (within the selected NUMA node). The proposed framework is flexible enough to accommodate any multi-objective criterion, while it is appropriately designed to handle noisy observations.

We demonstrated the utility of the proposed framework in the maximization of the running average processing speed of the threads and we evaluated its performance in four benchmark parallel applications. We have concluded that the \RL\ scheduler can achieve better running speed in certain cases, especially of small availability of resources or large background load. These observations should be further reinforced with additional benchmark tests. In addition, we plan to identify and generalize the indicators that trigger these advantageous responses of the \RL\ scheduler and also to consider additional utility functions, such as register count of each thread. 



\bibliographystyle{IEEEtran}
\bibliography{./bibliography}

\end{document}